# Formation of one-dimensional self-assembled silicon nanoribbons on Au(110)-(2x1)

Mohamed Rachid Tchalala[1], Hanna Enriquez[1], Andrew J. Mayne[1], Abdelkader Kara[2], Silvan Roth[3], Mathieu G. Silly[4], Azzedine Bendounan[4], Fausto Sirotti[4], Thomas Greber[3], Bernard Aufray[5], Gérald Dujardin[1], Mustapha Ait Ali[6] and Hamid Oughaddou[1,7,*]

[1]*Institut des Sciences Moléculaires d'Orsay, ISMO-CNRS, Bât. 210, Université Paris-Sud, F-91405 Orsay, France*
[2]*Department of Physics, University of Central Florida, Orlando, FL 32816, USA*
[3]*Physik-Institut der Universität Zürich, Winterthurerstrasse 190, CH-8057 Zürich, Switzerland*
[4]*TEMPO Beamline, Synchrotron Soleil, L'Orme des Merisiers Saint-Aubin, B.P. 48, F-91192 Gif-sur-Yvette Cedex, France*
[5]*CINaM-CNRS, Campus de Luminy, Marseille Cedex 09, F-13288 France*
[6] *Laboratoire de Chimie de Coordination et Catalyse, Département de Chimie, Faculté des Sciences-Semlalia, Université Cadi Ayyad, Marrakech, 40001, Morocco*
[7]*Département de Physique, Université de Cergy-Pontoise, F-95031 Cergy-Pontoise Cedex, France*

*Corresponding author: Prof. H. Oughaddou

Abstract:

We report results on the self-assembly of silicon nanoribbons on the (2x1) reconstructed Au(110) surface under ultra-high vacuum conditions. Upon adsorption of 0.2 monolayer (ML) of silicon the (2x1) reconstruction of Au(110) is replaced by an ordered surface alloy. Above this coverage a new superstructure is revealed by low electron energy diffraction (LEED) which becomes sharper at 0.3 Si ML. This superstructure corresponds to Si nanoribbons all oriented along the $[\bar{1}10]$ direction as revealed by LEED and scanning tunneling microscopy (STM). STM and high-resolution photoemission spectroscopy indicate that the nanoribbons are flat and predominantly 1.6 nm wide. In addition the silicon atoms show signatures of two chemical environments corresponding to the edge and center of the ribbons.

The remarkable properties of graphene have encouraged scientists to explore the feasibility of silicene, the analogue of graphene for silicon atoms[1-5]. Calculations have shown that silicene has an intrinsic stability and presents electronic properties similar to those of graphene[5]. From the experimental point of view, silicene was grown first on Ag(110)[6-9] and then on Ag(111) surfaces[10-13]. Deposition of silicon on a Ag(110) surface produces 1.6 nm wide silicene nanoribbons (NRs) with a honeycomb structure[7]. Evidence of a strong metallic character in these NRs was provided by high-resolution photoemission spectroscopy (HR-PES) measurements[6]. These data show quantized states in the valence band presenting one-dimensional (1D) dispersion along the NRs[9]. In addition, the oxidation of the NRs was shown to be less efficient compared to the oxidation of silicon surface[14]. On the Ag(111) surface, the growth of a continuous 2D sheet of silicene with large areas of an almost defect free honeycomb structure presenting a $(2\sqrt{3}\times2\sqrt{3})$ R30° superstructure, has been observed by STM[11]. Theoretical simulations (DFT / GGA) performed on this system confirm the stability of such a structure[7]. Recently, several experimental studies have independently reported the growth of silicene on the Ag(111) surface, revealing the existence of different ordered phases[12-13]. On both the (111) and (110) faces, the different phases are related to the intrinsic perfect match between four nearest neighbor Ag-Ag distances (1.156 nm) and three unit cells of the (111) surface of silicon (1.152 nm).

We have studied the growth of Si on Au[15]. Indeed Au is also suitable substrate because the Si-Au system presents a tendency to phase separation like Si-Ag[16]. In addition, the lattice parameter of Au (0.409 nm) is very close to that of Ag (0.408 nm) and the Si surface energy (1.200 J/m$^2$) is smaller than that of Au (1.506 J/m$^2$)[17]. From this point of view we can expect the formation of silicene NRs on top of the Au(110) surface similar to those on Ag(110).

Recently, the formation of a 2D crystal (Au–Si alloy) floating on top of the eutectic liquid has been reported[18-20]. During the first stages of silicon deposition on Au(110)-(2x1), we also observed a new ordered 2D surface alloy when the substrate is held at the eutectic temperature (~ 400°C)[15]. In this letter we present the growth characteristics of silicon NRs on Au(110) as well as their electronic properties. We show that exposure of the ordered Au-Si alloy surface layer to silicon results in the growth of parallel assembly of straight silicon NRs. The NRs have the same orientation and a predominantly characteristic narrow width of 1.6 nm, with lengths reaching a few hundreds of nanometers, limited by the characteristic Au(110) terrace width.

The experiments were performed under ultra-high vacuum (UHV) using the standard tools for surface preparation and characterization: an ion gun for surface cleaning, a low energy electron diffractometer (LEED) for structural characterization, a STM for surface characterization at the atomic scale working at room temperature (RT), and an Auger electron spectrometer (AES) for chemical surface analysis and calibration of the silicon coverage. The Au(110) sample was cleaned by several cycles of sputtering (600 eV Ar$^+$ ions, P = 5 x 10$^{-5}$ mbar) and annealing at 450°C until a sharp p(2x1) LEED pattern was obtained. Silicon, evaporated by direct current heating of a piece of Si wafer, was deposited onto the Au(110) surface held at a temperature of 400°C. The temperature was controlled by a thermocouple located close to the sample. The same crystal was studied in ISMO-Orsay for the LEED/STM experiments[21] and at the University of Zurich for detailed LEED analysis[22]. The high-resolution photoemission spectroscopy (HR-PES) angle-integrated measurements of the Si *2p* and Au *4f* core-levels were performed at RT on the TEMPO beam-line of the synchrotron SOLEIL[23].

Figure 1a shows the (2x1) LEED pattern characteristic of the bare Au(110) surface reconstruction. During silicon deposition on the Au(110) substrate held at 400°C, the (2x1) starts to disappear while a new superstructure appears at a Si coverage of ~ 0.2 monolayer (ML) which was shown to correspond to an ordered 2D surface alloy[15]. Above this coverage the diffraction spots corresponding to the 2D surface alloy disappear, and a rectangular superstructure appears, becoming sharp at 0.3 Si ML. The corresponding LEED pattern is shown in Figure 1b. The x6 periodicity is very clear along the [$\bar{1}10$]* while the spots along [001]* direction are not very distinct showing different spacing, suggesting different short-range order along this direction.

The STM images were recorded after the sample has been cooled down to RT. Figure 2a shows an atomically resolved filled-state STM image of the bare Au(110)-(2x1) surface. The missing-row structure of the Au(110)-(2x1) reconstruction is clearly observed. Figure 2b displays the STM topography of the silicon NRs formed after the deposition of 0.3ML silicon at 400°C. These NRs are all perfectly aligned along the [$\bar{1}10$] direction of the Au(110) surface. Only the surface steps limit the length of these nanoribbons and they have a low density of defects. In Figure 3a, the periodicity of 1.76 nm along the [$\bar{1}10$] direction (see also the line profile presented in Figure 3b) matches the periodicity of 6 x $a_1$ = 1.734 nm ($a_1$ = 0.289 nm) seen in the LEED pattern of Figure 1b. The x6 periodicity indicates that the NRs are not made of Si [$\bar{1}10$] rows with a "bulk-like" inter-atomic distance. In such a case a x4 periodicity would be expected, given the excellent match expected between four Au atomic distances (4 x 0.289 nm = 1.156 nm) and three silicon distances along Si [$\bar{1}10$] (3 x 0.385 nm = 1.155 nm). The line profile presented in Figure 3c shows that these NRs have a width of 1.6 nm, which corresponds to 4 x $a_2$ = 1.636 nm ($a_2$ = 0.409 nm) atomic distances along the [001] Au direction and a maximum apparent height of 0.24 nm (atomic height). In addition, the height profile of Figure 3-c also shows that the NRs are asymmetric across their widths.

Figure 4-a and Figure 4-b show the Au *4f* core level recorded on the pristine Au(110)-(2x1) surface at normal and at surface sensitive emission (75° with respect to the substrate normal), respectively, as well as their de-convolutions into two spin–orbit split components (B) and (S) located at 84.01 eV and 83.65 eV binding energies, respectively. A polynomial type secondary electron background is used and all spectra are fitted with a Doniach-Sunjic line shape[24]. To de-convolute the Au *4f* core level of the pristine surface, the following parameters were used: Gaussian widths of 115 meV and 150 meV for the component (B) and (S), respectively, and a 320 meV Lorentzian width with an asymmetry parameter equal to 0.02. The spin-orbit splitting was 3.678 eV and the branching ratio was 0.66. Because its intensity increases in the surface sensitive emission mode, the (S) component is attributed to the surface Au atoms while the (B) component corresponds to Au atoms in a bulk environment. The surface-atom component is located 0.36 eV below the bulk binding energy, which is in good agreement with the value reported in previous works[25-27].

Figures 4-c and 4-d show the Au *4f* core level spectra, recorded at normal and at surface sensitive emission, respectively, after growing Si NRs on the surface (0.3 Si ML). Two spin–orbit split components (B) and (S') are necessary to fit the spectra, with the (B) component still located at the same binding energy of 84.01 eV as the bulk component of the Au substrate (Figure 4a). However, the component (S') is now located at a binding energy of 84.60 eV, compared to 83.65 eV for the bare Au(110)-(2x1) surface. This component is now located at an energy 0.6 eV above the bulk binding energy. The Gaussian widths used to fit the components (B) and (S') are 215 meV and 415 meV, respectively, while the Lorentzian width and the asymmetry parameters are the same as in the pristine Au(110)-(2x1) surface. We also observe that the intensity of the component (S') increases in the surface sensitive emission mode, which indicates that it corresponds to the surface Au atoms with a Si environment. It is known that the (2x1) reconstruction is induced by the strain derivative of surface energy in the non-reconstructed Au(110) surface[28]. On the other hand, adsorbed atoms can affect the surface reconstruction of metallic (110) surfaces through charging because of the difference in the atomic work functions[29]. Our experiment shows that the (2x1) reconstruction of the Au(110) surface disappears upon Si adsorption. This explain why the component (S) vanishes and the component (S') appears. We have shown previously that the (2x1) reconstruction of Au(110) is replaced by an ordered surface alloy[16]. After the formation of the Si NRs, the LEED experiments do not show the (2x1) reconstruction.

The 2p core levels spectra of the adsorbed silicon are shown in Figures 5-a and 5-b. The Si *2p* spectra of the Si NRs are recorded at normal and surface sensitive emission together with their de-convolutions into two spin–orbit split components (S1) and (S2) located at 99.76 eV and 99.58 eV, respectively. These indicate two Si environments within the Si NRs. The fitting parameters used are: 150 meV Gaussian width for the (S1) component and 215 meV for the (S2) component; and an 80 meV Lorentzian width. The spin-orbit splitting is 0.605 eV and the branching ratio is 0.52. The best fit is obtained after including an asymmetry parameter of 0.09, which is higher than the 0.02 asymmetry parameter for the Au *4f* core levels. This is evidence for metallicity of the Si NRs. The very narrow total widths of the components, 130 meV for (S1) and 210 meV for (S2) indicate a high ordering of the Si atoms within the NRs. The Si *2p* core-levels spectra measured at normal emission and at surface sensitive emission are almost identical, with the same ratio between the two components. This indicates that all Si atoms are located on the surface and at a similar height in agreement with the STM images of the Si NRs.

We observe that the Si NRs obtained on Au(110)-(2x1) present similarities with those obtained on the Ag(110) surface[6-9]. Both are metallic and present protrusions within the NRs larger than one atom size, they are 0.16 nm wide, and they are asymmetric across their width. The HR-PES study indicates that, in both cases, Si atoms are present in two different environments. We have shown in the high resolution STM images of NRs on Ag(110), a honeycomb structure which is evidence for silicene[7].

However, in the present experimental study the atomic resolution of the honeycombs for Si/Au(110) is not seen in the STM topographies. This might be related to electron quantum interference at the NR edges as seen in graphene[30]. The STM images obtained routinely on Ag(110) also present protrusions larger than one atom size[9] suggesting similar behavior. Hence we deduced that the Si NRs formed on Au(110)-(2x1) could be silicene NRs.

It should be noted that complex Au/Si structures have been reported, such as the 6x6 structure of Au/Si(111)[31] and also the formation of an ordered surface alloy on Au(110)[15]. For the nano-ribbon phase reported here, considerations of the coincidence with the lattice of silicene allow the proposition of a simple ribbon model as indicated in Figure 6b. Along the close packed [-110] direction 6 Au atomic distances fit on 5 silicene unit cells with a lattice constant of 0.346 nm. Along the [100] direction, the ribbon width of about 4 Au lattice constants corresponds well with the width of 4 silicene rows. The Si-Si in-plane nearest neighbor lateral distance deduced from this model is 0.2 nm close to the value found for a silicene sheet on Ag(111)[11]. If there is some buckling as it was shown for silicene on Ag (111)[5-32], the apparent lateral Si-Si distance here of 0.2 nm will be shorter than the real Si-Si bond length. The model presents an asymmetry across the width of the Si NR in good agreement with the STM image (Figure 6-a). In addition the model shows that the silicon NRs are composed of five zig-zag chains of Si atoms running the length of the ribbon. Two zig-zag chains (one on each side of the ribbon) compose the edge of the NR, resulting in a fraction of 0.4 for the edge atoms in the unit cell. If we assign the (S1) component to the center atoms and the (S2) component to the edge atoms, their relative intensities are respectively 0.52 and 0.48 (measured) in qualitative agreement with the respective 0.6 and 0.4 ratios (from the model).

In conclusion, we have grown silicon NRs on Au(110) that are 1.6 nm wide with a periodicity x6 along the NR, as indicated by STM images and much similar to those obtained on Ag(110), in earlier experiments. High-resolution photoemission spectroscopy data on the silicon core levels indicate the presence of Si atoms in two environments (edge and middle atoms).In addition, our analysis also indicates that all Si atoms occupy the same height above the Au surface with a clear metallic character, hinting that silicene NRs on Au(110), have been grown. Atomistic calculations in progress will clarify the structure of these Si NRs.

**Acknowledgements**

The authors wish to thank Dr A.P. Seitsonen for fruitful discussions. RT and HO acknowledge the financial support from the European Community FP7-ITN Marie-Curie Programme (LASSIE project, grant agreement #238258) and the help of Prof. J.L. Lemaire its scientific coordinator in France at the Observatoire de Paris. Beam time was allocated under the Soleil project 20120189.

**Figures captions:**

**Figure 1:** LEED pattern recorded at Ep = 40eV corresponding to: a) the (2x1) reconstruction of the bare Au(110) surface (black rectangle), b) the x6 superstructure is visible inside the 2x1 rectangle after the deposition of ~ 0.3 Si ML on Au(110) held at 400°C.

**Figure 2:** a) atomically resolved filled-states STM image of the clean Au(110)-(2x1) reconstruction (4x4 nm$^2$, V = -80 mV, I = 2.2 nA). b) STM image of the surface at 0.3 Si ML coverage showing NRs oriented along the $[\bar{1}10]$ direction (60 x 60 nm$^2$, V = -0.2 V, I = 1.9 nA).

**Figure 3:** a) High resolution STM image showing the structure of the 1.6 nm wide Si NRs. (8 x 8 nm$^2$, V = -64 mV, I = 3.5 nA). b) Line profile along A showing the x6 periodicity. c) Line profile along B showing the asymmetry across the width of the NR.

**Figure 4:** Au 4f core levels spectra (dots) and their de-convolutions (solid line overlapping the data points) with two asymmetric components (B) and (S) recorded at hv=147eV for (a and b) the pristine Au(110)-(2x1) and (c and d) after deposition of 0.3 Si ML. The spectra (a) and (c) are recorded in normal emission while the spectra (b) and (d) are recorded in at 75° off-normal. The fitting parameters applied to the spectra of the pristine Au(110)-(2x1) are described in the main text. The (S) and (B) components are attributed to the surface and bulk Au atoms respectively of the pristine Au(110)-(2x1) while the (S') component corresponds to Au atoms bonded with Si atoms.

**Figure 5:** Si 2p core levels spectra (dots) and their de-convolutions (solid line overlapping the data points) with two asymmetric components (S1) and (S2) recorded after deposition of 0.3 Si ML on Au(110)-(2x1) at hv=147eV in (a) normal emission and (b) 75° off-normal. The fitting parameters are described in the main text. The (S1) and (S2) components are assigned to the center and the edge Si atoms respectively.

**Figure 6:** a) STM image of a Si NR showing the internal order with the asymmetry across its width. b) Proposed model based on the STM image and the LEED pattern. The Si NR with zig-zag edges is presented by a black honeycomb lattice. The atoms of the 1$^{st}$ and the 2$^{nd}$ Au layer are in light gray and dark gray respectively. The asymmetry across the width of the NR is highlighted. The Au atoms corresponding to the corners of the lozenge are in blue.

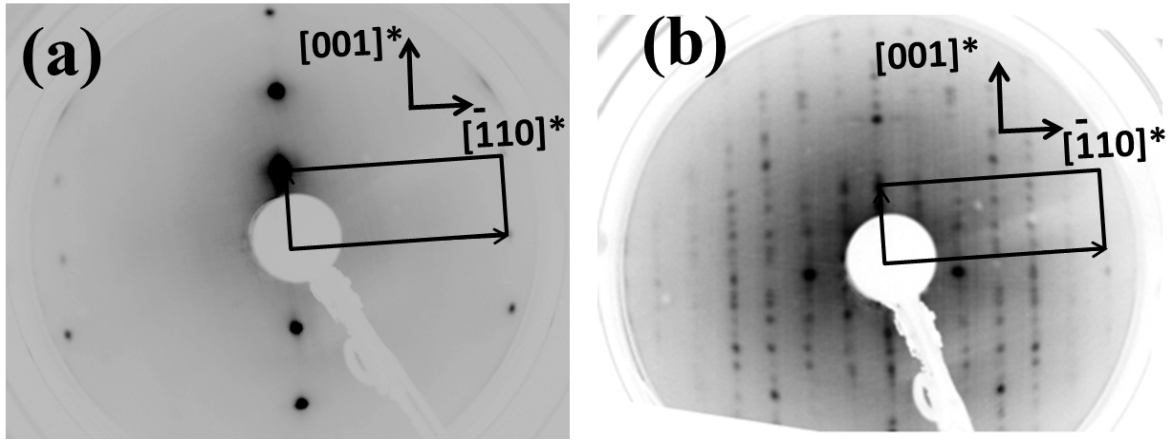

**Figure 1**

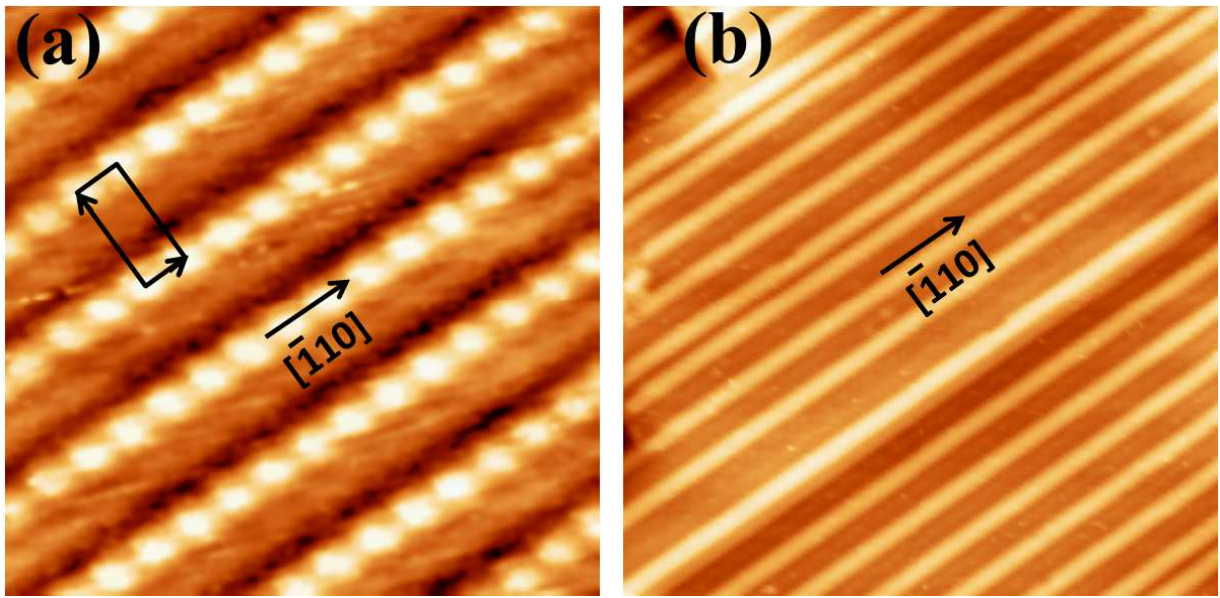

**Figure 2**

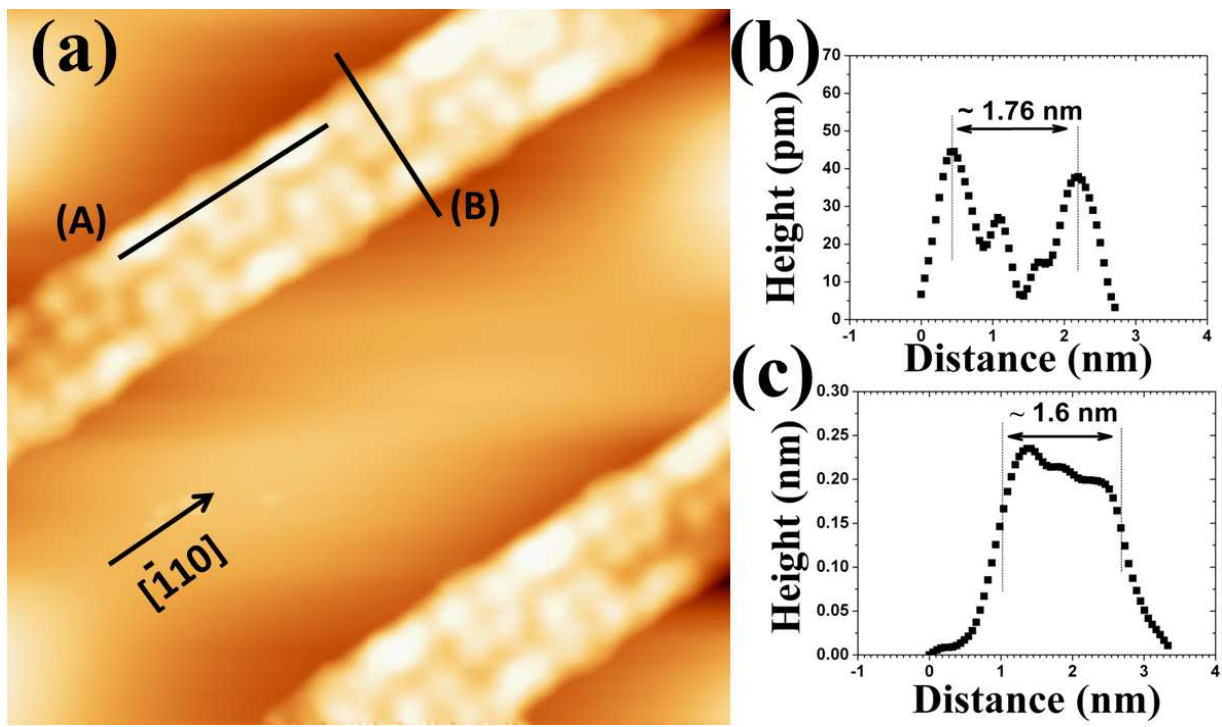

**Figure 3**

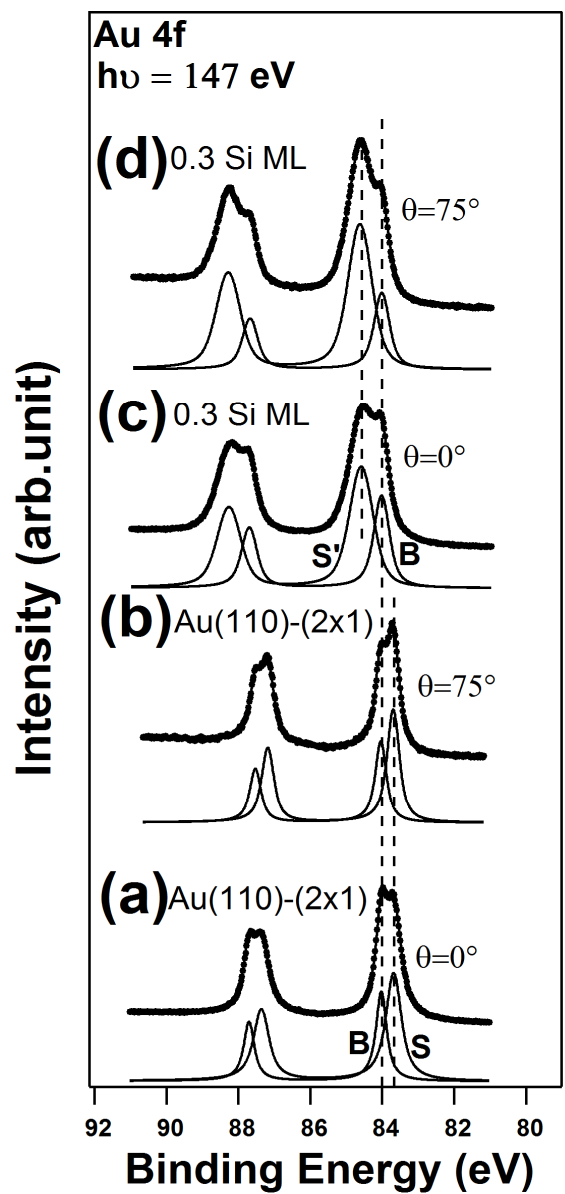

Figure 4

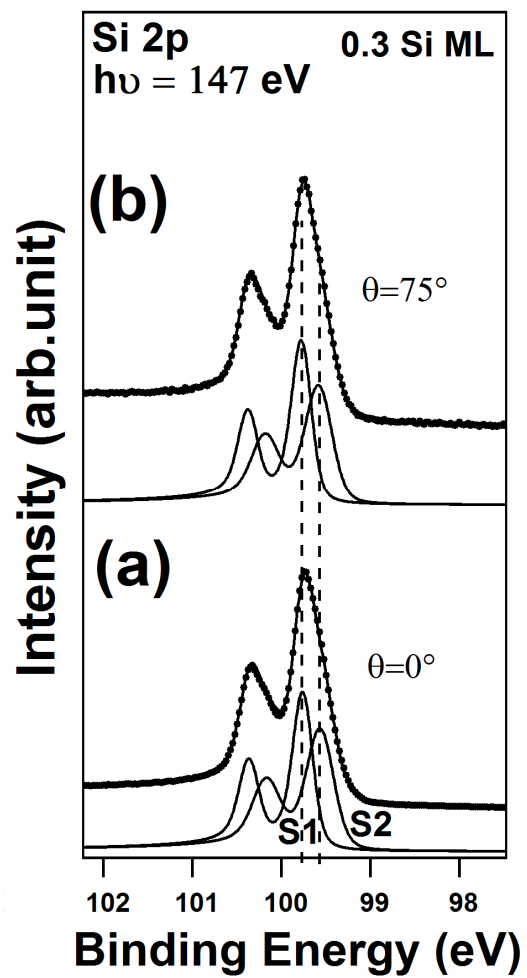

Figure 5

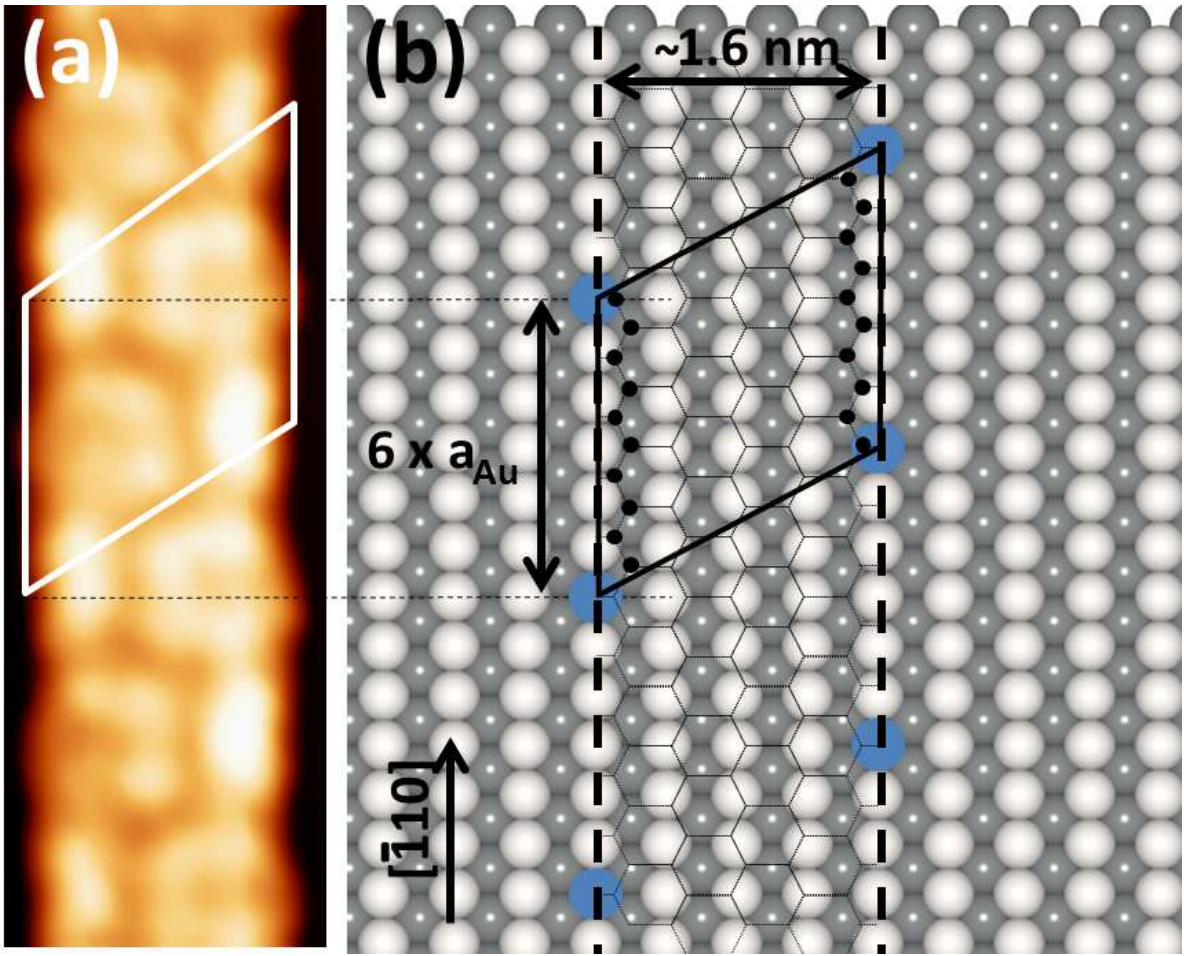

**Figure 6**